\documentclass[11pt,a4paper]{article}
\pdfoutput=1
\usepackage{jheppub}
\usepackage{slashed}

\makeatletter
\def\@fpheader{\relax}
\makeatother

\usepackage{ulem}

\usepackage[czech,english]{babel}
\usepackage{graphicx}
\usepackage{amsmath,amsfonts,amssymb}
\usepackage{url}
\usepackage{mathabx}
\DeclareMathOperator{\MyProd}{\scalebox{1.4}{$\mathrm{I\kern-0.2ex I}$}}

\preprint{LCTP-20-03}

\title{Toward  an Effective CFT$_2$ from ${\cal N}=4$ Super Yang-Mills and Aspects of Hawking Radiation }

\author[a]{Jun Nian}

\emailAdd{nian@umich.edu}

\affiliation[a]{Leinweber Center for Theoretical Physics, University of Michigan, Ann Arbor, MI 48109, U.S.A.}

\author[a, b]{and Leopoldo A. Pando Zayas}

\emailAdd{lpandoz@umich.edu}

\affiliation[b]{The Abdus Salam International Centre for Theoretical Physics, 34014 Trieste, Italy}

\abstract{Using ${\cal N}=4$ supersymmetric Yang-Mills theory we recover important aspects of the near-extremal thermodynamics of AdS$_5$ black holes including both the outer and the inner horizons with their corresponding entropy and energy. This $\mathcal{N}=4$ supersymmetric Yang-Mills theory  approach to black hole thermodynamics leads to an effective CFT$_2$ interpretation similar to the work by Callan and Maldacena. We corroborate this effective CFT$_2$ by implementing a particular near-horizon limit that geometrizes the Virasoro algebras as asymptotic symmetries. Using the effective CFT$_2$  picture, we discuss aspects of the Hawking radiation rate for a region of the near-extremal AdS$_5$ black hole quantum evolution.}


\arxivnumber{}

\newcommand{\bea}{\begin{eqnarray}}
\newcommand{\eea}{\end{eqnarray}}

\newcommand{\be}{\begin{equation}}
\newcommand{\ee}{\end{equation}}

\begin{document}

\maketitle

\section{Introduction}\label{sec:Introduction}

Hawking established the legitimacy of associating a temperature to black holes by explicitly demonstrating that due to quantum effects they radiate thermally \cite{Hawking:1974sw}.  This development cemented the interpretation of black hole entropy proportional to horizon area \cite{Bekenstein:1973ur} and, more generally, the thermodynamic nature of black holes \cite{Bardeen:1973gs, Bekenstein:1974ax, Hawking:1976de}. The search for a statistical description of black hole thermodynamics in terms of microstates became paramount. String theory, in the works of Strominger and Vafa provided such an answer for a particular class of supersymmetric asymptotically flat black holes  \cite{Strominger:1996sh}. In this case an understanding of the rate of Hawking radiation was provided by Callan and Maldacena who identified an effective two-dimensional conformal field theory for the microscopic description  \cite{Callan:1996dv}. Namely, they considered a near-extremal black hole configuration whose microscopic description involves a CFT with different temperatures  in  the right- and left-moving sectors  \cite{Callan:1996dv}. Other effective CFT$_2$ descriptions were discussed at the time \cite{Das:1996ug,Horowitz:1996fn,Das:1996wn} and perhaps the apex of this approach was provided by Strominger who presented a microscopic derivation of the entropy of a large class of black holes using only properties of a near-horizon limit \cite{Strominger:1997eq}.

It is natural to attempt to match such understanding in the framework of the AdS/CFT correspondence  \cite{Maldacena:1997re}. For AdS$_5$ black holes  the solutions have been constructed over a decade ago \cite{Cvetic:2004hs, Cvetic:2004ny, Chong:2005da, Chong:2005hr, Wu:2011gq}, the thermodynamical properties related to the inner and outer horizons have been discussed \cite{Cvetic:1996kv, Larsen:1997ge, Cvetic:1997uw, Cvetic:1997xv, Cvetic:1997vp, Larsen:1999pp, Cvetic:2010mn, Castro:2012av, Cvetic:2018dqf}.  A number of recent works has provided microscopic foundations for the black hole entropy using the dual supersymmetric field theories \cite{Cabo-Bizet:2018ehj, Choi:2018hmj, Benini:2018ywd, Cabo-Bizet:2019eaf}. Similar computations have been carried out for other asymptotically AdS black holes \cite{Choi:2019miv, Choi:2019zpz, Kantor:2019lfo, Nahmgoong:2019hko, Nian:2019pxj}.  It is the right time to ask whether an effective explanation for these microscopic results can be derived.

In this paper we take some steps toward the construction of such a universal description of the microscopic foundations for asymptotically AdS black holes. Our picture provides a concrete path to the understanding of the rate of Hawking radiation for near-extremal asymptotically AdS$_5$ black holes from the dual $\mathcal{N}=4$ SYM, which generalizes the previous work by Callan and Maldacena \cite{Callan:1996dv} for asymptotically flat black holes and the work by Gubser, Klebanov and Peet on near-extremal black D3-branes \cite{Gubser:1996de}. We elaborate on the approach of  \cite{Larsen:2019oll} to near-extremal AdS$_5$ black holes and obtain the entropy and energy at the outer and the inner horizons both from the gravity side and from the dual $\mathcal{N}=4$ SYM. The results can be grouped into a left and a right sector. This left-right-structure signals an underlying effective CFT$_2$ and  suggests the possibility of a locally AdS$_3$ near-horizon geometry. We, indeed, find such region via a near-horizon scaling for rotating black holes proposed in   \cite{Bardeen:1999px}. By applying the Cardy formula to the Virasoro algebra from the near-horizon local AdS$_3$, we obtain the same result for the growth of states  as from $\mathcal{N}=4$ SYM. With this validation we discuss aspects of Hawking radiation   only as detected in the near-horizon region.

This paper is organized as follows. In Sec.~\ref{sec:AdS5BH} we review the non-extremal AdS$_5$ black hole solutions in the literature,  and systematically distinguish the supersymmetric, the extremal and the BPS conditions for AdS$_5$ black holes. We also discuss the near-extremal thermodynamical relations at the outer horizon, and then generalize these relations to the inner horizon. In Sec.~\ref{sec:N=4SYM} we give a microscopic derivation of the thermodynamical relations for near-extremal AdS$_5$ black holes from the boundary $\mathcal{N}=4$ SYM. In Sec.~\ref{sec:CFT2} we demonstrate that the results obtained in the previous sections can be formulated as emerged from an effective CFT$_2$, which can also be justified by taking a special near-horizon limit similar to the Kerr/CFT correspondence. In Sec.~\ref{sec:HawkingRadiation}, we apply these notions to the early-time near-horizon Hawking radition, and find that indeed the radiation rate is thermal. Some possible directions for the future research are discussed in Sec.~\ref{sec:Discussions}.

\section{Near-Extremal AdS$_5$ Black Hole Thermodynamics}\label{sec:AdS5BH}

Let us briefly review the non-extremal electrically charged rotating AdS$_5$ black holes, i.e. the non-extremal Kerr-Newman AdS$_5$ black holes. We follow closely \cite{Chong:2005hr}, which discusses black holes with generic angular momenta $J_1$ and $J_2$ but degenerate electric charges $Q_1 = Q_2 = Q_3 \equiv Q$ within minimal gauged supergravity, which generalizes the non-extremal neutral rotating Kerr AdS$_5$ black holes discussed in \cite{Hawking:1998kw} as well as some previously found special charged solutions \cite{Cvetic:2004hs, Cvetic:2004ny, Chong:2005da}. More general solutions with arbitrary $Q_I$'s were later constructed in \cite{Wu:2011gq}.

The bosonic part of the 5d minimal gauged supergravity is given by the Lagrangian:
\be\label{eq:SUGRA_Lagrangian}
 {\cal L} = (R + 12 g^2) * \mathbb{I} - \frac{1}{2} * F \wedge F + \frac{1}{3 \sqrt{3}} F \wedge F \wedge A\, ,
\ee
where $F = dA$, and $g = L^{-1} > 0$ with $L$ denoting the AdS$_5$ radius.

It was found in \cite{Chong:2005hr} that the equations of motion of the theory \eqref{eq:SUGRA_Lagrangian} have the following solution in the Boyer-Lindquist coordinates $x^\mu = (t,\, r,\, \theta,\, \phi,\, \psi)$:
\begin{align}
  ds^2 & = - \frac{\Delta_\theta \left[(1 + g^2 r^2) \rho^2 dt + 2 q \nu \right]\, dt}{\Xi_a \Xi_b \rho^2} + \frac{2 q \nu \omega}{\rho^2} + \frac{f}{\rho^4} \left(\frac{\Delta_\theta dt}{\Xi_a \Xi_b} - \omega \right)^2 + \frac{\rho^2 dr^2}{\Delta_r} + \frac{\rho^2 d\theta^2}{\Delta_\theta} \nonumber\\
  {} & \quad + \frac{r^2 + a^2}{\Xi_a}\, \textrm{sin}^2 \theta\, d\phi^2 + \frac{r^2 + b^2}{\Xi_b}\, \textrm{cos}^2 \theta\, d\psi^2\, ,\label{eq:AdS5metric}\\
  A & = \frac{\sqrt{3}\, q}{\rho^2} \left(\frac{\Delta_\theta\, dt}{\Xi_a \Xi_b} - \omega\right)\, ,
\end{align}
where
\begin{align}
\begin{split}
  \nu & \equiv b\, \textrm{sin}^2 \theta\, d\phi + a\, \textrm{cos}^2 \theta\, d\psi\, ,\\
  \omega & \equiv a\, \textrm{sin}^2 \theta\, \frac{d\phi}{\Xi_a} + b\, \textrm{cos}^2 \theta\, \frac{d\psi}{\Xi_b}\, ,\\
  \Delta_\theta & \equiv 1 - a^2 g^2\, \textrm{cos}^2 \theta - b^2 g^2\, \textrm{sin}^2 \theta\, ,\\
  \Delta_r & \equiv \frac{(r^2 + a^2) (r^2 + b^2) (1 + g^2 r^2) + q^2 + 2 a b q}{r^2} - 2 m\, ,\\
  \rho^2 & \equiv r^2 + a^2\, \textrm{cos}^2 \theta + b^2\, \textrm{sin}^2 \theta\, ,\\
  \Xi_a & \equiv 1 - a^2 g^2\, ,\\
  \Xi_b & \equiv 1 - b^2 g^2\, ,\\
  f & \equiv 2 m \rho^2 - q^2 + 2 a b q g^2 \rho^2\, .
\end{split}
\end{align}

The angular momenta $J_i$ and the electric charge $Q$ of the AdS$_5$ black hole can be obtained as follows. From the Komar integral one obtains the angular momenta:
\be
  J_i = \frac{1}{16 \pi} \int_{S^3} * dK_i\, ,
\ee
where $i = 1,\, 2$, and
\be
  K_1 = \frac{\partial}{\partial \phi}\, ,\quad K_2 = \frac{\partial}{\partial \psi}\, .
\ee
More explicitly,
\begin{align}
\begin{split}
  J_1 & = \frac{\pi \Big[2 a m + q b (1 + a^2 g^2) \Big]}{4 \Xi_a^2 \Xi_b}\, ,\\
  J_2 & = \frac{\pi \Big[2 b m + q a (1 + b^2 g^2) \Big]}{4 \Xi_b^2 \Xi_a}\, .
\end{split}
\end{align}
From the Gaussian integral one obtains the electric charge:
\be
  Q = \frac{1}{16 \pi} \int_{S^3} \left(* F - \frac{1}{\sqrt{3}} F \wedge A \right)\, .
\ee
More explicitly,
\be
  Q = \frac{\sqrt{3} \pi q}{4\, \Xi_a \Xi_b}\, .
\ee

To obtain the energy (or conserved mass) $E$, we distinguish the outer and the inner horizons. At the outer horizon, the energy is obtained by integrating the first law of thermodynamics:
\be
  dE = T_+ dS_+ + \Omega_i^+ dJ_i + \Phi^+ dQ\, ,
\ee
where $\Phi^+ \equiv \ell^\mu A_\mu$ is the electric potential at the outer horizon. The explicit expression of the energy is
\be\label{eq:Energy}
  E = \frac{m \pi (2 \Xi_a + 2 \Xi_b - \Xi_a \Xi_b) + 2 \pi q a b g^2 (\Xi_a + \Xi_b)}{4 \Xi_a^2 \Xi_b^2}\, .
\ee

At the inner horizon, we can obtain the angular momenta $J_i$, the electric charge $Q$ and the energy $E$ in the same way, but the first law of thermodynamics at the inner horizon looks different:
\be
  dE = - T_- dS_- + \Omega_i^- dJ_i + \Phi^- dQ\, .
\ee
A similar expression of the first law has been seen in the previous study on the asymptotically flat black holes \cite{Castro:2012av}. From a direct computation, we see that the expressions of the energy $E$, the angular momenta $J_i$ and the electric charge $Q$ remain the same at both the outer and the inner horizons.

When the AdS$_5$ black hole solution preserves some supersymmetries, the BPS bound is  saturated, i.e., the conserved quantities characterizing the black holes satisfy
\be
  E - g J_1 - g J_2 - 3 Q = 0\, ,
\ee
which is equivalent to the condition
\be\label{eq:GravityBPSconstr-1}
  q = \frac{m}{1 + a g + b g}\, .
\ee
More precisely, the black hole solutions with the supersymmetric condition \eqref{eq:GravityBPSconstr-1} have $\frac{1}{4}$-BPS supersymmetry.

One can also characterize the black hole based on the properties of the inner, $r_-$,  and outer, $r_+$, horizons.  We define the extremal solution as the one for which the outer and the inner horizons coincide, i.e.,  $r_+ = r_-$, and the temperature becomes zero, i.e. $T_+ = T_- = 0$. Hence, the extremal limit is achieved when $\Delta_r (r)$ has a double root, or equivalently, the discriminant of $\Delta_r (r)$ vanishes, which can be viewed as an equation for $m$. By solving this equation, we obtain the extremal value of $m$, i.e. $m_{\textrm{ext}}$, as a function of $(a, b, q)$
\be\label{eq:GravityExtcond}
  m = m_{\textrm{ext}} (a, b, q)\, .
\ee
Here, we omit the lengthy expression of $m_{\textrm{ext}} (a, b, q)$. When $m > m_{\textrm{ext}}$, the factor $\Delta_r$ has two different real roots corresponding to the outer and the inner horizons, while for $m < m_{\textrm{ext}}$ the factor $\Delta_r$ does not have any real roots, which indicates  a naked singularity visible from the asymptotic infinity, instead of a black hole solution. Similar conditions for AdS$_4$ black holes have been discussed in \cite{Caldarelli:1999xj, David:2020jhp}.

A BPS black hole solution satisfies both the supersymmetric condition \eqref{eq:GravityBPSconstr-1} and the extremal condition \eqref{eq:GravityExtcond}. Under the supersymmetric condition \eqref{eq:GravityBPSconstr-1}, the extremal value $m_{\textrm{ext}}$ can be expressed as
\be\label{eq:GravityBPSconstr-2}
  m_{\textrm{ext}} = \frac{1}{g} (a + b) (1 + a g) (1 + b g) (1 + a g + b g)\, .
\ee
which can also be obtained by requiring that the BPS black hole solutions have no unphysical naked closed timelike curves, as shown in \cite{Chong:2005hr}.

\begin{figure}[!htb]
\begin{center}
  \includegraphics[width=0.6\textwidth]{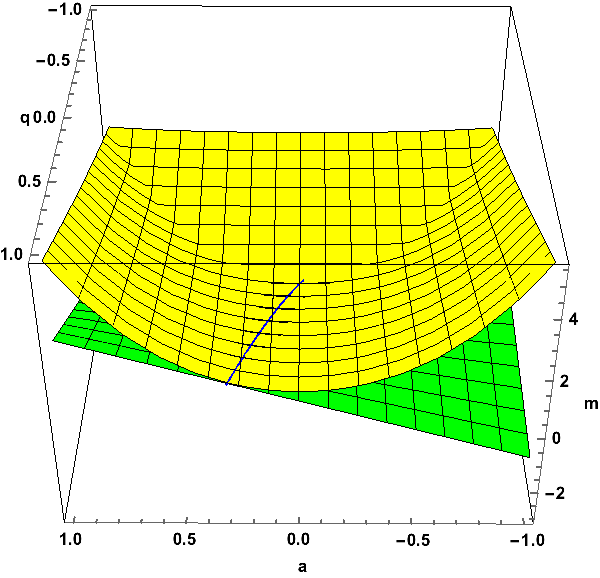}
  \caption{The extremal surface (yellow), the supersymmetric surface (green) and the BPS surface as their intersection (blue)}
   \label{fig:ParameterSurfaces}
\end{center}
\end{figure}

To illustrate the relation of the supersymmetric condition \eqref{eq:GravityBPSconstr-1} and the extremal condition \eqref{eq:GravityExtcond}, we plot them as two codimension-1 surfaces in the parameter space $(a, b, q, m)$ in Fig.~\ref{fig:ParameterSurfaces}, where we set $b = a$ for simplicity. The BPS black hole solutions correspond to the intersection of these two codimension-1 surfaces.

The positions of the horizons are  simplified in the supersymmetric limit: 
\be\label{eq:BPSRootEq}
  r^2 \Delta_r = \frac{1}{g^2} \left(a + b + a b g - g r^2 \right)^2 \Big[\left(1 + a g + b g \right)^2 + g^2 r^2 \Big] = 0\, .
\ee
The outer horizon $r_+$ and the inner horizon $r_-$ coincide in the BPS limit:
\be
  r_0^2 \equiv \frac{a + b + a b g}{g}\, .
\ee
Considering a black hole slightly away from the BPS limit, the degeneracy of $r_+$ and $r_-$ is lifted by a small change $0< \delta r \ll r_0$, i.e.,
\be\label{eq:TwoHorizons}
  r_+ \, \longrightarrow\, r_0 + \delta r\, ,\quad\quad r_- \, \longrightarrow\, r_0 - \delta r\, .
\ee

One can prove that the vector
\begin{align}
  \ell (r) & \equiv \frac{\partial}{\partial t} + \Omega_1 (r) \frac{\partial}{\partial \phi} + \Omega_2 (r) \frac{\partial}{\partial \psi}\\
\begin{split}
  \textrm{with}\quad \Omega_1 (r) & \equiv \frac{a (r^2 + b^2) (1 + g^2 r^2) + b q}{(r^2 + a^2) (r^2 + b^2) + a b q}\, ,\\
  \Omega_2 (r) & \equiv \frac{b (r^2 + a^2) (1 + g^2 r^2) + a q}{(r^2 + a^2) (r^2 + b^2) + a b q}\, ,
\end{split}
\end{align}
becomes null, i.e. $g_{\mu \nu} \ell^\mu \ell^\nu = 0$, when the equation $\Delta_r = 0$ holds. Therefore, the vectors $\ell (r_\pm)$ are null Killing vectors at the outer horizon $r_+$ and the inner horizon $r_-$ respectively,  defining Killing horizons. From these Killing vectors,  $\ell (r_\pm)$, one can find the corresponding surface gravities $\kappa (r_\pm)$ obeying $\ell^\nu \nabla_\nu \ell^\mu = \kappa \ell^\mu$ in the Eddington-Finkelstein coordinates, which are proportional to $\Delta'_r (r_\pm) / 2$, respectively. The temperatures associated with the horizons can be defined by the surface gravity $\kappa$ at $r_\pm$:
\be\label{eq:temperature}
  T_\pm = \Bigg|\frac{\kappa (r_\pm)}{2 \pi} \Bigg| = \frac{r_\pm^4 \Big[1 + g^2 (2 r_\pm^2 + a^2 + b^2) \Big] - (a b + q)^2}{2 \pi r_\pm \Big[(r_\pm^2 + a^2) (r_\pm^2 + b^2) + a b q \Big]}\, .
\ee
The entropies at the horizons are proportional to the surface area of the horizons:
\be\label{eq:SBH from Gravity}
  S_\pm = \frac{\pi^2 \Big[(r_\pm^2 + a^2) (r_\pm^2 + b^2) + a b q \Big]}{2 \Xi_a \Xi_b r_\pm}\, .
\ee

At the inner horizon we can either choose a positive temperature \cite{Castro:2012av} or a negative temperature \cite{Cvetic:2018dqf}. We choose a positive temperature at the inner horizon in this paper. With this choice, the first laws of thermodynamics look different at two horizons. At the outer horizon,
\be\label{eq:FirstLawOuterHorizon}
  dE = T_+ dS_+ + \Omega_i^+ dJ_i + \Phi^+ dQ\, ,
\ee
while at the inner horizon,
\be\label{eq:FirstLawInnerHorizon}
  dE = - T_- dS_- + \Omega_i^- dJ_i + \Phi^- dQ\, ,
\ee
which differ by a sign.

Let us focus on the near-extremal case in this paper. For such configurations, we should treat the thermodynamics at the outer and the inner horizons separately. As we have seen in \eqref{eq:FirstLawOuterHorizon} and \eqref{eq:FirstLawInnerHorizon}, the first laws at the two horizons differ by a sign, similar to asymptotically flat black holes \cite{Castro:2012av}. At the outer horizon, the following relations have been obtained from the near-extremal Kerr-Newman AdS$_5$ black holes in \cite{Larsen:2019oll}:
\be\label{eq:EntropyEnergyOuterHorizon}
  S_+ = S_* + \left(\frac{C}{T} \right)_*\, T_+\, ,\quad E = E_* + \frac{1}{2} \left(\frac{C}{T} \right)_*\, T_+^2\, ,
\ee
\be\label{eq:C over T from gravity}
  \left(\frac{C}{T} \right)_* = \frac{\pi^2 g^{-1} \Big[8 (Q / g)^3 + \frac{1}{4} N^4 (J_1 + J_2) \Big]}{3 (Q / g)^2-\frac{1}{2} N^2 \left(J_1 + J_2 \right) + \left(3 Q / g + \frac{1}{2} N^2\right)^2}\, ,
\ee
with $Q$ and $J_i$ taking the values for the BPS black holes.  Above,  $(C/T)_*$ is the heat capacity from the outer horizon on the gravity side  which provides a counting for near extremal microstates.

\section{Microscopic ${\cal{N}}=4$ SYM Approach}\label{sec:N=4SYM}

\subsection{Microstate Counting of BPS AdS$_5$ Black Holes} 
It has recently been shown in \cite{Cabo-Bizet:2018ehj, Choi:2018hmj, Benini:2018ywd}, that entropy of electrically charged, rotating BPS AdS$_5$ black holes can be obtained from $\mathcal{N}=4$ SYM by computing the superconformal index with complex chemical potentials, $\Delta_I$:
\be
  \mathcal{I} = \textrm{Tr} \Big[(-1)^F e^{-\beta E} e^{- \Delta_I Q_I - \omega_i J_i} \Big]\, .
\ee
In the Cardy limit, for small angular velocites, $|\omega_i| \ll 1$, the leading results of the superconformal index is:
\be\label{eq:BPSPartFct}
  \mathcal{F} \equiv \textrm{log}\, Z \simeq \textrm{log}\, \mathcal{I} \simeq \frac{N^2}{2} \frac{\Delta_1 \Delta_2 \Delta_3}{\omega_1 \omega_2}\, ,
\ee
subject to a constraint with $s_I, t_i \in \{\pm 1\}$:
\be
  \sum_I s_I\, \Delta_I - \sum_i t_i\, \omega_i = 2 \pi i\, .
\ee

A more detailed discussion of the superconformal index in the context of ${\cal N}=4$ SYM was presented in \cite{Honda:2019cio,ArabiArdehali:2019tdm}. The leading expression for the superconformal index  has been obtained for generic ${\cal N}=1$ supersymmetric field theories exploiting a Cardy-like limit in  \cite{Cabo-Bizet:2019osg,Kim:2019yrz,Amariti:2019mgp}. In two other works \cite{Lezcano:2019pae,Lanir:2019abx}, the leading contribution to the superconformal index of generic ${\cal N}=1$ supersymmetric theories was computed without recourse to any Cardy limit. This was facilitated by using the Bethe-Ansatz formulation of the index \cite{Benini:2018mlo, Benini:2018ywd} which is based on the original work of \cite{Closset:2017bse}. A novel approach to the index, emphasizing modular properties, has recently been introduced in  \cite{Cabo-Bizet:2019eaf} and makes contact with aspects of the Bethe-Ansatz approach.

Let us briefly summarize the extremization process of the entropy function for the BPS AdS$_5$ black holes \cite{Cabo-Bizet:2018ehj, Choi:2018hmj, Benini:2018ywd}. After the extremization, one obtains the entropy of $\frac{1}{16}$-BPS electrically charged rotating AdS$_5$ black holes. We follow closely \cite{Cabo-Bizet:2018ehj, Larsen:2019oll}.

First, from the leading order of $\mathcal{N}=4$ SYM's partition function, one can define the entropy function via a Legendre transformation:
\be
  S (\Delta_I,\, J_i,\, \Lambda) = - \frac{N^2}{2} \frac{\Delta_1 \Delta_2 \Delta_3}{\omega_1 \omega_2} - Q_I \Delta_I - J_i\, \omega_i - \Lambda \Big( \sum_I \Delta_I - \sum_i \omega_i - 2 \pi i \Big)\, .\label{eq:EntropyFct}
\ee
where $\Lambda$ is a Lagrange multiplier that imposes one choice of the BPS constraint $\sum_I s_I \Delta_I - \sum_i t_i \omega_i = 2 \pi i$ with $s_I = t_i = +1$. The entropy function can be viewed as a functional of the electric potentials $\Delta_I$ and the angular momenta $J_i$.

Extremizing the entropy function \eqref{eq:EntropyFct} with respect to $\Delta_I$ and $\omega_i$, we obtain the equations
\begin{align}
  \frac{\partial S}{\partial \Delta_I} & = - \frac{N^2}{2} \frac{\Delta_1 \Delta_2 \Delta_3}{\Delta_I \omega_1 \omega_2} - (Q_I + \Lambda) = 0\, ,\label{eq:BPSExtEq-1}\\
  \frac{\partial S}{\partial \omega_i} & = \frac{N^2}{2} \frac{\Delta_1 \Delta_2 \Delta_3}{\omega_i\, \omega_1 \omega_2} - (J_i - \Lambda) = 0\, .\label{eq:BPSExtEq-2}
\end{align}
Combining the equations \eqref{eq:BPSExtEq-1} and \eqref{eq:BPSExtEq-2}, we obtain
\be\label{eq:CubicLambda}
  \prod_I (Q_I + \Lambda) + \frac{N^2}{2} \prod_i (J_i - \Lambda) = 0\, ,
\ee
which is a cubic equation for the Lagrange multiplier $\Lambda$. Moreover, simplifying the entropy function \eqref{eq:EntropyFct} using the extremization conditions \eqref{eq:BPSExtEq-1} and \eqref{eq:BPSExtEq-2}, we find the BPS black hole entropy
\be\label{eq:EntropyLambda}
  S_* = 2 \pi i \Lambda\, .
\ee
In order to express the black hole entropy in terms of $Q_I$ and $J_i$, we need to solve the cubic equation \eqref{eq:CubicLambda} to express $\Lambda$ as a function of $Q_I$ and $J_i$. We first write \eqref{eq:CubicLambda} as
\begin{align}
  & \Lambda^3 + \Lambda^2  \left(Q_1 + Q_2 + Q_3 + \frac{1}{2} N^2 \right) + \Lambda  \left(Q_1 Q_2 + Q_2 Q_3 + Q_3 Q_1 - \frac{1}{2} N^2 (J_1 + J_2) \right) \nonumber\\
  & + \left(Q_1 Q_2 Q_3 + \frac{1}{2} N^2 J_1 J_2 \right) = 0\, ,
\end{align}
or equivalently,
\be\label{eq:CubicExplicit-1}
  \Lambda^3 + A \Lambda^2 + B \Lambda + C = 0
\ee
with
\begin{align}
\begin{split}
  A & = Q_1 + Q_2 + Q_3 + \frac{1}{2} N^2\, ,\\
  B & = Q_1 Q_2 + Q_2 Q_3 + Q_3 Q_1 - \frac{1}{2} N^2 (J_1 + J_2)\, ,\\
  C & = Q_1 Q_2 Q_3 + \frac{1}{2} N^2 J_1 J_2\, .
\end{split}
\end{align}
By requiring that the entropy is real, \eqref{eq:EntropyLambda} implies that $\Lambda$ has a purely imaginary root. Since $Q_I$ and $J_i$ are all real, the purely imaginary roots for $\Lambda$ must appear in pairs for the cubic equation \eqref{eq:CubicExplicit-1}. Consequently, the cubic equation \eqref{eq:CubicExplicit-1} should take the form 
\be\label{eq:CubicExplicit-2}
(\Lambda^2 + B) (\Lambda + A) = \Lambda^3 + A \Lambda^2 + B \Lambda + A B =0\, .
\ee
Comparing \eqref{eq:CubicExplicit-1} and \eqref{eq:CubicExplicit-2}, we therefore obtain the constraint $C = A B$, more explicitly,
\begin{align}\label{eq:ChargeConstraint}
  & \Bigg(Q_1 + Q_2 + Q_3 + \frac{N^2}{2} \Bigg)\,  \Bigg(Q_1 Q_2 + Q_2 Q_3 + Q_3 Q_1 - \frac{N^2}{2} (J_1 + J_2) \Bigg)\nonumber\\
  & - \Bigg(Q_1 Q_2 Q_3 + \frac{N^2}{2} J_1 J_2\Bigg) = 0\, . 
\end{align}
This is the constraint on $Q_I$ and $J_i$ for the BPS AdS$_5$ black holes. For generic values of $Q_I$ and $J_i$ we define a function as $h \equiv C - A B$. Hence, for the BPS case the constrait \eqref{eq:ChargeConstraint} becomes $h=0$.

Subject to the constraint \eqref{eq:ChargeConstraint}, the purely imaginary root $\Lambda = - i \sqrt{B}$ gives the physical entropy for the $\frac{1}{16}$-BPS electrically charged rotating AdS$_5$ black holes
\be\label{eq:S_BPS}
  S = 2\pi i \Lambda = 2\pi \sqrt{Q_1 Q_2 + Q_2 Q_3 + Q_3 Q_1 - \frac{N^2}{2} (J_1 + J_2) }\, .
\ee
This also agrees with the result on the gravity side \cite{Kim:2006he}. Corresponding to the purely imaginary root $\Lambda = - i \sqrt{B}$, the solutions to the extremization equations \eqref{eq:BPSExtEq-1} \eqref{eq:BPSExtEq-2} are
\begin{align}
  \frac{\Delta_I}{2 \pi i} & = \frac{\prod_K (Q_K + \Lambda)}{Q_I + \Lambda} \frac{1}{2 \Lambda (\Lambda + Q_1 + Q_2 + Q_3 + \frac{1}{2} N^2)}\, ,\label{eq:SolDeltaI}\\
  \frac{\omega_i}{2 \pi i} & = \frac{N^2}{2} \frac{\prod_k (J_k - \Lambda)}{J_i - \Lambda} \frac{1}{2 \Lambda (\Lambda + Q_1 + Q_2 + Q_3 + \frac{1}{2} N^2)}\, .\label{eq:Solomegai}
\end{align}

\subsection{Inner and Outer Horizons} 

In this subsection we discuss the extremization process for the near-BPS AdS$_5$ black holes, following closely \cite{Larsen:2019oll}. To study the near-BPS AdS$_5$ black holes, we distinguish two classes: the near-extremal ones and the extremal near-BPS ones. The near-extremal configurations can be obtained by introducing a small temperature $T$, while the extremal near-BPS ones are characterized by the violation $\varphi$  of the BPS constraint $\sum_I s_I \Delta_I - \sum_i t_i \omega_i = 2 \pi i$.

Let us first consider the near-extremal configurations. Usually in field theory, the temperature is introduced as $T = \beta^{-1}$, where $\beta$ denotes the period of the Euclidean time circle $S^1$. However, the $\frac{1}{16}$-BPS index of $\mathcal{N}=4$ SYM does not depend on $\beta$, hence, $\beta$ cannot be simply interpreted as $T^{-1}$ in this case. Instead, we make use of the following relations to connect the temperature derivatives of the quantities on the gravity side, $\Phi_I$ and $\Omega_i$, with the real parts of the quantities on the field theory side, $\Delta_I$ and $\omega_i$:
\be\label{eq:BPStoNearExtRelations}
  \textrm{Re} (\Delta_I) = \frac{\partial \Phi_I}{\partial T}\, ,\quad \textrm{Re} (\omega_i) = \frac{\partial \Omega_i}{\partial T}\, .
\ee
These relations were found in \cite{Cabo-Bizet:2018ehj} and checked explicitly in \cite{Larsen:2019oll}.

If we assume that there is no phase transition  between  BPS black holes and near-extremal ones, that is, that the number of degrees of freedoms varies smoothly with a small temperature, then the spectrum of the microstates in the dual field theory should also change smoothly from BPS to near-extremal ones. The self-consistency of such crucial assumption will be verified  {\it a posteriori}  by the fact that there is no jump in the heat capacity from $T=0$ to an infinitesimal $T>0$. Based on this assumption, we can use the relations \eqref{eq:BPStoNearExtRelations} to extend the BPS partition function \eqref{eq:BPSPartFct} of the $\mathcal{N}=4$ SYM to the one at an infinitesimal temperature $T > 0$ \cite{Larsen:2019oll}:
\be
  \textrm{log}\, Z = \frac{N^2}{2 T} \frac{(\Phi_1 - \Phi_1^*) (\Phi_2 - \Phi_2^*) (\Phi_3 - \Phi_3^*)}{(\Omega_1 - \Omega_1^*) (\Omega_2 - \Omega_2^*)}\, ,
\ee
and it is related to other thermodynamic quantities in the following way:
\be
  \textrm{log}\, Z = S - \frac{1}{T} (M - M_*) - \frac{1}{T} (\Phi_I - \Phi_I^*) Q_I - \frac{1}{T} (\Omega_i - \Omega_i^*) J_i\, .
\ee
Moreover, $(\Phi_I - \Phi_I^*)$ and $(\Omega_i - \Omega_i^*)$ satisfy an additional constraint:
\be\label{eq:ConstrNearBPS}
  \sum_I (\Phi_I - \Phi_I^*) - \sum_i (\Omega_i - \Omega_i^*) = \varphi + 2 \pi i T\, ,
\ee
where $\varphi$ denotes the violation of the BPS constraint $\sum_I s_I \Delta_I - \sum_i t_i \omega_i = 2 \pi i$, that characterizes the extremal near-BPS case, while the near-extremal case is parameterized by $T$. In the BPS limit, both $\varphi$ and $T$ vanish, hence, $\Phi_I$ and $\Omega_i$ will approach their BPS values $\Phi_I^*$ and $\Omega_i^*$ respectively.

We can define a free energy for the near-BPS AdS$_5$ black holes as the counterpart of the BPS entropy function \eqref{eq:EntropyFct}:
\begin{align}
  \mathcal{F} & \equiv (M - M_*) - T S - \Lambda \Bigg[\sum_I (\Phi_I - \Phi_I^*) - \sum_i (\Omega_i - \Omega_i^*) - \varphi - 2 \pi i T \Bigg]\\
  {} & = - \frac{N^2}{2} \frac{(\Phi_1 - \Phi_1^*) (\Phi_2 - \Phi_2^*) (\Phi_3 - \Phi_3^*)}{(\Omega_1 - \Omega_1^*) (\Omega_2 - \Omega_2^*)} - (\Phi_I - \Phi_I^*) Q_I - (\Omega_i - \Omega_i^*) J_i \nonumber\\
  {} & \quad - \Lambda \Bigg[\sum_I (\Phi_I - \Phi_I^*) - \sum_i (\Omega_i - \Omega_i^*) - \varphi - 2 \pi i T \Bigg]\, ,\label{eq:FreeEnergyFct}
\end{align}
where $\Lambda$ is again a Lagrange multiplier that imposes the near-BPS constraint \eqref{eq:ConstrNearBPS}.

We see that the near-BPS free energy \eqref{eq:FreeEnergyFct} has an expression similar to the BPS entropy function \eqref{eq:EntropyFct}. Hence, we can perform a similar extremization process:
\begin{align}
  \frac{\partial S}{\partial (\Phi_I - \Phi_I^*)} & = - \frac{N^2}{2} \frac{(\Phi_1 - \Phi_1^*) (\Phi_2 - \Phi_2^*) (\Phi_3 - \Phi_3^*)}{(\Phi_I - \Phi_I^*) \omega_1 \omega_2} - (Q_I + \Lambda) = 0\, ,\label{eq:NearBPSExtEq-1}\\
  \frac{\partial S}{\partial (\Omega_i - \Omega_i^*)} & = \frac{N^2}{2} \frac{(\Phi_1 - \Phi_1^*) (\Phi_2 - \Phi_2^*) (\Phi_3 - \Phi_3^*)}{(\Omega_i - \Omega_i^*) (\Omega_1 - \Omega_1^*) (\Omega_2 - \Omega_2^*)} - (J_i - \Lambda) = 0\, .\label{eq:NearBPSExtEq-2}
\end{align}
These extremization equations have the solutions:
\begin{align}
  \frac{\Phi_I - \Phi_I^*}{\varphi + 2 \pi i T} & = \frac{\prod_K (Q_K + \Lambda)}{Q_I + \Lambda} \frac{1}{2 \Lambda (\Lambda + Q_1 + Q_2 + Q_3 + \frac{1}{2} N^2)}\, ,\label{eq:SolPhiI}\\
  \frac{\Omega_i - \Omega_i^*}{\varphi + 2 \pi i T} & = \frac{N^2}{2} \frac{\prod_k (J_k - \Lambda)}{J_i - \Lambda} \frac{1}{2 \Lambda (\Lambda + Q_1 + Q_2 + Q_3 + \frac{1}{2} N^2)}\, ,\label{eq:SolOmegai}
\end{align}
which are similar to the solutions \eqref{eq:SolDeltaI} \eqref{eq:Solomegai} in the BPS case.

The extremal near-BPS case has been analyzed in more detail in \cite{Larsen:2019oll}, where it is shown that the following relations at the leading order can be obtained from the dual field theory:
\begin{align}
  S - S_* & = \left(\frac{C_E}{T} \right)_* \left(\frac{\varphi}{2 \pi} \right)\, ,\\
  E - E_* & = \frac{1}{2} \left(\frac{C_T}{T} \right)_* \left(\frac{\varphi}{2 \pi} \right)^2\, ,
\end{align}
where
\begin{align}
  \left(\frac{C_T}{T} \right)_* & = \, \frac{\pi^2 \Bigg[ (Q_1 + Q_2) (Q_2 + Q_3) (Q_3 + Q_1) + \frac{1}{4} N^4 (J_1 + J_2) \Bigg]}{Q_1 Q_2 + Q_2 Q_3 + Q_3 Q_1 - \frac{1}{2} N^2 (J_1 + J_2) + \left(Q_1 + Q_2 + Q_3 + \frac{1}{2} N^2 \right)^2}\, ,\\
  \left(\frac{C_E}{T} \right)_* & = \frac{2 \pi}{S_*} \left(\frac{C_T}{T} \right)_* \left(Q_1 + Q_2 + Q_3 + \frac{1}{2} N^2 \right)\, .
\end{align}
These results are consistent with the ones obtained on the gravity side. However, the near-extremal case has not been considered from the dual field theory side in \cite{Larsen:2019oll}.

To derive the relations \eqref{eq:EntropyEnergyOuterHorizon} for the near-extremal case from $\mathcal{N}=4$ SYM, the BPS cubic equation \eqref{eq:CubicExplicit-2} gets modified:
\be\label{eq:LambdaEqNearExt-1}
  \Big[\left(\Lambda + \delta \Lambda \right)^2 + B \Big] \Big[ (\Lambda + \delta \Lambda) + A \Big] + h = 0\, ,
\ee
with $h \neq 0$ for the near-extremal case. It means that a change in $\Lambda$, denoted by $\delta \Lambda$, will cause a small violation of the BPS constraint, $h_* = 0$. Comparing \eqref{eq:CubicExplicit-2} and \eqref{eq:LambdaEqNearExt-1}, we obtain at the leading order in $\delta \Lambda$:
\be\label{eq:LambdaEqNearExt-2}
  (3 \Lambda^2 + 2 A \Lambda + B) \delta \Lambda + h = 0\, .
\ee
Since we consider the near-extremal case by perturbing the BPS case, whose black hole entropy corresponds to one of the purely imaginary roots satisfying $\Lambda^2 + B = 0$, the equation \eqref{eq:LambdaEqNearExt-2} can be simplified. Since only the imaginary part of $\delta \Lambda$ contributes to the real part of $\delta S$, assuming that $h$ is purely imaginary, we obtain
\be
  \textrm{Im} (\delta \Lambda) = - h\, \textrm{Re} \Bigg[\frac{1}{2 \Lambda A + 2 \Lambda^2} \Bigg] = \frac{h}{2 (A^2 - \Lambda^2)} = \frac{h}{2 (A^2 + B)}\, .
\ee
Consequently, the change of the black hole entropy is
\be\label{eq:NearExtEntropy}
  \delta S \equiv S_+ - S_* = 2 \pi i\, \delta \Lambda = \frac{i \pi h}{A^2 + B}\, .
\ee
To find the entropy excess of the near-extremal black holes compared to the BPS ones, we should relate the function $h$ in \eqref{eq:NearExtEntropy} to the temperatures $T_\pm$, similar to the extremal near-BPS case discussed in \cite{Larsen:2019oll}.  

The change $\delta h$ can be obtained from
\be
  \delta h = \frac{\partial h}{\partial Q_I} \delta Q_I + \frac{\partial h}{\partial J_i} \delta J_i\, ,
\ee
by introducing the transformation parameter $\lambda$, as:
\be
  \delta Q_I = \lambda Q_I\, ,\quad \delta J_i = \lambda J_i\, ,\quad 2 \pi i\, \delta T_+ = 2 \lambda\, .\label{eq:dT+}
\ee
By varying $h = C - A B$ we obtain
\begin{align}
  \delta h & = - \Bigg[(Q_1 + Q_2) (Q_2 + Q_3) (Q_3 + Q_1) + \frac{1}{4} N^4 (J_1 + J_2) \Bigg] \lambda\, ,\nonumber\\
  {} & = - \pi i\, \Bigg[(Q_1 + Q_2) (Q_2 + Q_3) (Q_3 + Q_1) + \frac{1}{4} N^4 (J_1 + J_2) \Bigg]\, \delta T_+\, ,\label{eq:Rel dh and dT}
\end{align}
which is simplified using the BPS constraint $h_* = 0$. Since for the near-extremal black holes away from the BPS ones ($T_* = 0$, $h_* = 0$) we have $\delta h = h$ and $\delta T_+ = T_+$, \eqref{eq:Rel dh and dT} leads to
\be\label{eq:Rel h and T - 1}
  h = - \pi i\, \Bigg[ (Q_1 + Q_2) (Q_2 + Q_3) (Q_3 + Q_1) + \frac{1}{4} N^4 (J_1 + J_2) \Bigg] T_+\, ,
\ee
where $T_+$ denotes the temperature corresponding to the outer horizon. In the degenerate limit $Q_1 = Q_2 = Q_3 \equiv Q$, combining \eqref{eq:NearExtEntropy} and \eqref{eq:Rel h and T - 1}, we obtain at the order $\mathcal{O} (T_+)$:
\be\label{eq:C over T from bdy}
  S_+ - S_* = \frac{\pi^2 \Big[8 Q^3 + \frac{1}{4} N^4 (J_1 + J_2) \Big]}{3 Q^2 - \frac{1}{2} N^2 (J_1 + J_2) + \left(3 Q + \frac{1}{2} N^2 \right)^2}\, T_+ \equiv \left(\frac{C}{T} \right)_* T_+\, .
\ee
Integrating the first law at the outer horizon \eqref{eq:FirstLawOuterHorizon} with fixed values of $Q$ and $J_i$, we obtain
\be
  E - E_* = \frac{1}{2} \left(\frac{C}{T} \right)_* T_+^2\quad \textrm{with}\quad E_* \equiv J_1 + J_2 + 3 Q\, .
\ee
These results exactly match the ones \eqref{eq:EntropyEnergyOuterHorizon} \eqref{eq:C over T from gravity} derived on the gravity side.

We can similarly use $\mathcal{N}=4$ SYM to derive the entropy and the energy relations at the inner horizon. However, \eqref{eq:TwoHorizons} implies that $\delta T_- \propto - \delta r < 0$. Accoding to our choice \eqref{eq:temperature} of positive $T_-$ at the inner horizon, there should be $\delta T_- = - T_-$. Therefore, the relation between $h$ and $T_-$ has a sign change compared to the one between $h$ and $T_+$ \eqref{eq:Rel h and T - 1} and consequently,
\be
  S_- - S_* = - \left(\frac{C}{T} \right)_* T_-\, ,\quad E - E_* = \frac{1}{2} \left(\frac{C}{T} \right)_* T_-^2\, ,
\ee
where for the second equation we integrate the first law at the inner horizon \eqref{eq:FirstLawInnerHorizon} with fixed values of $Q$ and $J_i$.  To our knowledge, these results with an explicit heat capacity at the inner horizon have not appeared in the literature before, and we provide the first derivation from $\mathcal{N} = 4$ SYM.

\section{An  Effective CFT$_2$}\label{sec:CFT2}

\subsection{An Effective CFT$_2$ Near Extremality} 
Let us now reorganize the results obtained for $\mathcal{N}=4$ SYM in a way that suggets an underlying CFT$_2$ structure.

For the near-extremal case, $T_\pm$ coincide at the leading order. To uncover the existence of a left sector and a right sector, we define
\be\label{eq:Def SL and SR}
  S_L \equiv S_*\, ,\quad S_R \equiv \left(\frac{C}{T} \right)_*\, T_+ = \left(\frac{C}{T} \right)_*\, T_- + \mathcal{O} (\delta r^2)\, ,
\ee
\begin{align}
  T_L & \equiv \frac{(2 E_* / S_* + T_+) + (2 E_* / S_* - T_-)}{2} = \frac{2 E_*}{S_*} + \mathcal{O} (\delta r^2)\, ,\label{eq:DefTL}\\
  T_R & \sim \frac{(2 E_* / S_* + T_+) - (2 E_* / S_* - T_-)}{2} = T_+ + \mathcal{O} (\delta r^2)\, .\label{eq:DefTR}
\end{align}
Consequently, 
\be
  S_\pm = S_L \pm S_R = \frac{\pi^2}{3} c_L T_L \pm \frac{\pi^2}{3} c_R T_R\, ,\label{eq:S+ in c and T}
\ee 
where $c_{L} \equiv 3 S_*^2 / (2 \pi^2 E_*)$ and $c_{R} \sim (C / T)_*$ denote the central charges of the left and the right sectors respectively, both of which are of the order $\mathcal{O} (N^2)$. These relations are consistent with the ones for the asymptotically flat black holes \cite{Cvetic:1996kv, Larsen:1997ge, Cvetic:1997uw, Cvetic:1997xv, Cvetic:1997vp, Larsen:1999pp, Cvetic:2010mn, Castro:2012av}. If the temperatures and the mode numbers are related in the following way: 
\be\label{eq:Rel T and N}
  T_L = \frac{1}{\pi} \sqrt{\frac{6 N_L}{c_L}}\, ,\quad T_R = \frac{1}{\pi} \sqrt{\frac{6 N_R}{c_R}}\, ,
\ee
the expressions \eqref{eq:S+ in c and T} can be rewritten as
\be
  S_\pm = 2 \pi \sqrt{\frac{c_L N_L}{6}} \pm 2 \pi \sqrt{\frac{c_R N_R}{6}}\, .
\ee 
Equation \eqref{eq:Rel T and N} will be supported by a different analysis.

\subsection{Near-Horizon Metric}

The effective appearance of the left- and the right-moving modes in $\mathcal{N}=4$ SYM indicates the possibility of a geometric interpretation, which we have indeed found. In this subsection, we discuss the near-horizon metrics of AdS$_5$ black holes in several different coordinates.

As we can see in Fig.~\ref{fig:ParameterSurfaces}, when perturbing around the asymptotically AdS$_5$ $\frac{1}{4}$-BPS black holes, we can deviate from the extremal surface by increasing the parameter $m$, while keeping the other parameters $(a, b, q)$ fixed. More precisely, we relax the condition \eqref{eq:GravityBPSconstr-2} and expand the parameter $m$ around its BPS value $m_0$ for near-extremal AdS$_5$ black holes
\be\label{eq:AdS5 NearBPScond}
  m = m_0 (1 + \lambda^2\, \widetilde{m})\, ,
\ee
with
\be
  m_0 \equiv \frac{1}{g} (a + b) (1 + a g) (1 + b g) (1 + a g + b g)\, .
\ee
A similar near-extremal condition for AdS$_4$ black holes has been discussed in \cite{David:2020jhp}. In  what follows, we consider near-extremal AdS$_5$ black holes by imposing the condition \eqref{eq:AdS5 NearBPScond} and fixing the parameter $q$ to its BPS value $m_0 / (1 + a g + b g)$ given by \eqref{eq:GravityBPSconstr-1}. For simplicity, we also set $b = a$, hence, the black hole solutions are characterized by only one independent parameter $a$.

In addition, let us consider a near-horizon limit of the metric \eqref{eq:AdS5metric} of the non-extremal AdS$_5$ black hole following the scaling introduced by Bardeen and Horowitz in \cite{Bardeen:1999px}:
\be\label{eq:NearHorizonLimit}
  r \to r_0 + \lambda\, \widetilde{r}\, ,\quad t \to \frac{\widetilde{t}}{\lambda}\, ,\quad \phi \to \widetilde{\phi} + g \frac{\widetilde{t}}{\lambda}\, ,\quad \psi \to \widetilde{\psi} + g \frac{\widetilde{t}}{\lambda}\, .
\ee
In principle, for the near-horizon region of a non-extremal black hole we should expand $r$ around the outer horizon $r_+$ instead of the BPS black hole horizon $r_0$. However, since under the near-extremal condition \eqref{eq:AdS5 NearBPScond} $r_+$ and $r_0$ differ by a constant of order $\lambda$, we can absorb that constant into $\widetilde{r}$ and still expand $r$ around $r_0$.

By taking the limit $\lambda \to 0$, we find the following near-horizon metric:
\begin{align}
  ds^2 & = - \frac{\widetilde{\Delta}_{\textrm{AdS$_5$}} (\widetilde{r})}{a (1 + a g)^2} d\widetilde{t}^2 + \frac{a\, d\widetilde{r}^2}{\widetilde{\Delta}_{\textrm{AdS$_5$}} (\widetilde{r})} + \frac{2 a}{g (1 - a g)}\, d\theta^2 \nonumber\\
  {} & + \Lambda_{\textrm{AdS$_5$}} (\theta) \left[d\widetilde{\phi} + \frac{3 g (1 - a g)}{(1 + a g) \sqrt{a \left(a + \frac{2}{g} \right)}} \, \widetilde{r}\, d\widetilde{t} \right]^2 \nonumber\\
  {} & \quad + \frac{a \Big(4 - a g + 3 a g\, \textrm{cos} (2 \theta) \Big)\, \textrm{cos}^2 \theta}{2 g (1 - a g)^2} \left[d\widetilde{\psi} + \frac{6 a g\, \textrm{sin}^2 \theta}{4 - a g + 3 a g\, \textrm{cos} (2 \theta)} d\widetilde{\phi} + V_{\textrm{AdS$_5$}} (\theta)\, \widetilde{r}\, d\widetilde{t} \right]^2 \, ,\label{eq:AdS5 NH metric Global}
\end{align}
where
\begin{align}
\begin{split}\label{eq:DefAdS5Factors}
  \widetilde{\Delta}_{\textrm{AdS$_5$}} (\widetilde{r}) & \equiv 2 g (1 + 5 a g) \widetilde{r}^2 - 2\, a\, \widetilde{m}\, (1 + a g) (1 + 2 a g)\, ,\\
  \Lambda_{\textrm{AdS$_5$}} (\theta) & \equiv \frac{4 a (2 + a g)\, \textrm{sin}^2 \theta}{g (1 - a g) \Big(4 - a g + 3 a g\, \textrm{cos} (2 \theta) \Big)}\, ,\\
  V_{\textrm{AdS$_5$}} (\theta) & \equiv \frac{6 g (1 - a g) \sqrt{g \left(\frac{2}{a} + g\right)}}{(1 + a g) \left(4 - a g + 3 a g\, \textrm{cos} (2 \theta) \right)}\, .
\end{split}
\end{align}
Defining some new coordinates
\be\label{eq:Defrhat}
  \hat{r} \equiv \widetilde{r}\, \sqrt{\frac{g (1 + 5 a g)}{a (1 + a g) (1 + 2 a g) \widetilde{m}}}\, ,\quad \hat{t} \equiv 2 \widetilde{t}\, \sqrt{\frac{g (1 + 2 a g) (1 + 5 a g) \widetilde{m}}{a (1 + a g)}}\, ,
\ee
we can rewrite the near-horizon metric \eqref{eq:AdS5 NH metric Global} as
\begin{align}
  ds^2 & = \frac{a}{2 g (1 + 5 a g)} \bigg[- (\hat{r}^2 - 1) d\hat{t}^2 + \frac{d\hat{r}^2}{\hat{r}^2 - 1} \bigg] + \frac{2 a}{g (1 - a g)}\, d\theta^2 \nonumber\\
  {} & \quad + \Lambda_{\textrm{AdS$_5$}} (\theta) \left[d\widetilde{\phi} + \frac{3 a (1 - a g)}{2 (1 + 5 a g) \sqrt{a \left(a + \frac{2}{g} \right)}} \, \hat{r}\, d\hat{t} \right]^2 \nonumber\\
  {} & \quad + \frac{a \Big(4 - a g + 3 a g\, \textrm{cos} (2 \theta) \Big)\, \textrm{cos}^2 \theta}{2 g (1 - a g)^2} \left[d\widetilde{\psi} + \frac{6 a g\, \textrm{sin}^2 \theta}{4 - a g + 3 a g\, \textrm{cos} (2 \theta)} d\widetilde{\phi} + \hat{V}_{\textrm{AdS$_5$}} (\theta)\, \hat{r}\, d\hat{t} \right]^2 \, ,
\end{align}
where
\be\label{eq:DefVhat}
  \hat{V}_{\textrm{AdS$_5$}} (\theta) \equiv \frac{3 a (1 - a g) \sqrt{g \left(\frac{2}{a} + g\right)}}{(1 + 5 a g) \left(4 - a g + 3 a g\, \textrm{cos} (2 \theta) \right)}\, .
\ee
Let us introduce a further change of coordinates
\be\label{eq:PoincareTo Global 1}
  g\, \rho = \hat{r} + \sqrt{\hat{r}^2 - 1}\, \textrm{cosh} (\hat{t})\, ,\quad g^{-1}\, \tau = \frac{\sqrt{\hat{r}^2 - 1}\, \textrm{sinh} (\hat{t})}{\hat{r} + \sqrt{\hat{r}^2 - 1}\, \textrm{cosh} (\hat{t})}\, ,
\ee
which implies that
\begin{align}
\begin{split}\label{eq:PoincareTo Global 2}
  - \rho^2\, d\tau^2 + \frac{d\rho^2}{\rho^2} & = - (\hat{r}^2 - 1)\, d\hat{t}^2 + \frac{d\hat{r}^2}{\hat{r}^2 - 1}\, ,\\
  \rho\, d\tau & = \hat{r}\, d\hat{t} + d\Theta\, ,
\end{split}
\end{align}
with
\be\label{eq:PoincareTo Global 3}
  \Theta \equiv \textrm{log} \left(\frac{1 + \sqrt{\hat{r}^2 - 1}\, \textrm{sinh} (\hat{t})}{\textrm{cosh} (\hat{t}) + \hat{r}\, \textrm{sinh} (\hat{t})} \right)\, .
\ee 
This change of coordinates can bring the global coordinates $(\hat{t},\, \hat{r})$ into the Poincar\'e coordinates $(\tau,\, \rho)$. Consequently, the near-horizon metric \eqref{eq:AdS5 NH metric Global} becomes
\begin{align}
  ds^2 & = \frac{a}{2 g (1 + 5 a g)} \bigg[- \rho^2\, d\tau^2 + \frac{d\rho^2}{\rho^2} \bigg] + \frac{2 a}{g (1 - a g)}\, d\theta^2 \nonumber\\
  {} & \quad + \Lambda_{\textrm{AdS$_5$}} (\theta) \left[d\hat{\phi} + \frac{3 a (1 - a g)}{2 (1 + 5 a g) \sqrt{a \left(a + \frac{2}{g} \right)}} \, \rho\, d\tau \right]^2 \nonumber\\
  {} & \quad + \frac{a \Big(4 - a g + 3 a g\, \textrm{cos} (2 \theta) \Big)\, \textrm{cos}^2 \theta}{2 g (1 - a g)^2} \left[d\hat{\psi} + \frac{6 a g\, \textrm{sin}^2 \theta}{4 - a g + 3 a g\, \textrm{cos} (2 \theta)} d\hat{\phi} + \hat{V}_{\textrm{AdS$_5$}} (\theta)\, \rho\, d\tau \right]^2 \, ,\label{eq:AdS5 NHmetric}
\end{align}
where
\be\label{eq:Def phihat psihat}
  \hat{\phi} \equiv \widetilde{\phi} - \frac{3 a (1 - a g)}{2 (1 + 5 a g) \sqrt{a \left(a + \frac{2}{g} \right)}}\, \Theta\, ,\quad \hat{\psi} \equiv \widetilde{\psi} - \frac{3 a (1 - a g)}{2 (1 + 5 a g) \sqrt{a \left(a + \frac{2}{g} \right)}}\, \Theta\, .
\ee
We see that in the near-horizon metric \eqref{eq:AdS5 NHmetric} the coordinates $(\tau, \rho, \hat{\phi}, \hat{\psi})$ form two $U(1)$'s fibered over AdS$_2$, similar to the situation in the Kerr/CFT correspondence \cite{Guica:2008mu, Lu:2008jk}. Alternatively, the near-horizon metric \eqref{eq:AdS5 NHmetric} can also be written in the standard form:
\be\label{eq:AdS5 NHmetric standard form}
  ds^2 = f_0 (\theta) \left(- \rho^2\, d\tau^2 + \frac{d\rho^2}{\rho^2} \right) + f_\theta (\theta)\, d\theta^2 + \gamma_{ij} (\theta)\, \left(dx^i + k^i \rho\, d\tau \right)\, \left(dx^j + k^j \rho\, d\tau \right)
\ee
with $x^i \in \{\hat{\phi},\, \hat{\psi} \}$, and the coefficients $f_0 (\theta)$, $f_\theta (\theta)$, $\gamma_{ij} (\theta)$ and $k_i$'s can be easily obtained by comparing \eqref{eq:AdS5 NHmetric} with \eqref{eq:AdS5 NHmetric standard form}.

Besides the near-horizon Bardeen-Horowitz scaling \eqref{eq:NearHorizonLimit}, we can apply a different light-cone scaling in the near-horizon region introduced in \cite{Matsuo:2010ut}, which will be relevant for computing the right central charge $c_R$. To apply this near-horizon light-cone scaling, we first introduce the light-cone coordinates
\be
  x^+ \equiv \lambda \left(\psi + g t \right)\, ,\quad x^- \equiv \psi - g t\, ,
\ee
and then consider the near-horizon scaling in the new light-cone coordinates
\be
  r \to r_0 + \lambda\, \widetilde{r}\, ,\quad t \to \frac{x^+ - \lambda x^-}{2 g \lambda}\, ,\quad \phi \to \widetilde{\phi} + \frac{x^+ - \lambda x^-}{2 \lambda} \, ,\quad \psi \to \frac{x^+ + \lambda x^-}{2 \lambda}\, .
\ee
Together with the condition \eqref{eq:AdS5 NearBPScond} and taking the limit $\lambda \to 0$, we obatin the new near-horizon metric for the AdS$_5$ near-extremal black holes in the coordinates $(x^+,\, \widetilde{r},\, \theta, \widetilde{\phi},\, x^-)$
\begin{align}
  ds^2 & = - \frac{\widetilde{\Delta}_{\textrm{AdS$_5$}} (\widetilde{r})\, dx^{+2}}{4 a g^2 (1 + a g)^2} + \frac{a\, d\widetilde{r}^2}{\widetilde{\Delta}_{\textrm{AdS$_5$}} (\widetilde{r})} + \frac{2 a}{g (1 - a g)}\, d\theta^2 \nonumber\\
  {} & \quad + \Lambda_{\textrm{AdS$_5$}} (\theta) \left[d\widetilde{\phi} + \frac{3 (1 - a g)}{2 (1 + a g) \sqrt{a \left(a + \frac{2}{g} \right)}} \widetilde{r}\, dx^+ \right]^2 \nonumber\\
  {} & \quad + \frac{a \Big(4 - a g + 3 a g\, \textrm{cos} (2 \theta) \Big)\, \textrm{cos}^2 \theta}{2 g (1 - a g)^2} \left[dx^- + \frac{6 a g\, \textrm{sin}^2 \theta}{4 - a g + 3 a g\, \textrm{cos} (2 \theta)} d\widetilde{\phi} + \frac{V_{\textrm{AdS$_5$}} (\theta)}{2 g}\, \widetilde{r}\, dx^+ \right]^2\, ,\label{eq:AdS5 LightCone Metric}
\end{align}
where $\widetilde{\Delta}_{\textrm{AdS$_5$}} (\widetilde{r})$, $\Lambda_{\textrm{AdS$_5$}} (\theta)$ and $V_{\textrm{AdS$_5$}} (\theta)$ are the same as \eqref{eq:DefAdS5Factors}.
We can bring the metric in the light-cone coordinates \eqref{eq:AdS5 LightCone Metric} into an expression similar to the global coordinates \eqref{eq:AdS5 NH metric Global}.
\begin{align}
  ds^2 & = \frac{a}{2 g (1 + 5 a g)} \bigg[- (\hat{r}^2 - 1) d\widetilde{x}^{+2} + \frac{d\hat{r}^2}{\hat{r}^2 - 1} \bigg] + \frac{2 a}{g (1 - a g)}\, d\theta^2 \nonumber\\
  {} & \quad + \Lambda_{\textrm{AdS$_5$}} (\theta) \left[d\widetilde{\phi} + \frac{3 a (1 - a g)}{2 (1 + 5 a g) \sqrt{a \left(a + \frac{2}{g} \right)}} \, \hat{r}\, d\widetilde{x}^+ \right]^2 \nonumber\\
  {} & \quad + \frac{a \Big(4 - a g + 3 a g\, \textrm{cos} (2 \theta) \Big)\, \textrm{cos}^2 \theta}{2 g (1 - a g)^2} \left[dx^- + \frac{6 a g\, \textrm{sin}^2 \theta}{4 - a g + 3 a g\, \textrm{cos} (2 \theta)} d\widetilde{\phi} + \hat{V}_{\textrm{AdS$_5$}} (\theta)\, \hat{r}\, d\widetilde{x}^+ \right]^2 \, ,\label{eq:AdS5 LightCone Metric Global}
\end{align}
where $\hat{r}$ and $\hat{V}_{\textrm{AdS$_5$}} (\theta)$ are defined in the same way as \eqref{eq:Defrhat} and \eqref{eq:DefVhat} respectively, and
\be
  \widetilde{x}^+ \equiv x^+\, \sqrt{\frac{\widetilde{m} (1 + 2 a g) (1 + 5 a g)}{a g (1 + a g)}}\, .
\ee
Similar to \eqref{eq:PoincareTo Global 1} \eqref{eq:PoincareTo Global 2} \eqref{eq:PoincareTo Global 3} \eqref{eq:Def phihat psihat}, we can apply some further changes of coordinates to bring $(\widetilde{x}^+, \hat{r}, \theta, \widetilde{\phi}, x^-)$ to $(\hat{x}^+, \hat{\rho}, \theta, \hat{\phi}^+, \hat{x}^-)$
\be\label{eq:PoincareTo Global 4}
  g\, \hat{\rho} = \hat{r} + \sqrt{\hat{r}^2 - 1}\, \textrm{cosh} (\widetilde{x}^+)\, ,\quad g^{-1}\, \hat{x}^+ = \frac{\sqrt{\hat{r}^2 - 1}\, \textrm{sinh} (\widetilde{x}^+)}{\hat{r} + \sqrt{\hat{r}^2 - 1}\, \textrm{cosh} (\widetilde{x}^+)}\, ,
\ee
which leads to
\begin{align}
\begin{split}\label{eq:PoincareTo Global 5}
  - \hat{\rho}^2\, d\hat{x}^{+2} + \frac{d\hat{\rho}^2}{\hat{\rho}^2} & = - (\hat{r}^2 - 1)\, d\widetilde{x}^{+2} + \frac{d\hat{r}^2}{\hat{r}^2 - 1}\, ,\\
  \hat{\rho}\, d\hat{x}^+ & = \hat{r}\, d\widetilde{x}^+ + d\Theta^+\, ,
\end{split}
\end{align}
where
\be\label{eq:PoincareTo Global 6}
  \Theta^+ \equiv \textrm{log} \left(\frac{1 + \sqrt{\hat{r}^2 - 1}\, \textrm{sinh} (\widetilde{x}^+)}{\textrm{cosh} (\widetilde{x}^+) + \hat{r}\, \textrm{sinh} (\widetilde{x}^+)} \right)\, .
\ee
Consequently, the metric \eqref{eq:AdS5 LightCone Metric Global} can be written into an expression similar to the Poincar\'e coordinates \eqref{eq:AdS5 NHmetric}.

\begin{align}
  ds^2 & = \frac{a}{2 g (1 + 5 a g)} \bigg[- \hat{\rho}^2\, d\hat{x}^{+2} + \frac{d\hat{\rho}^2}{\hat{\rho}^2} \bigg] + \frac{2 a}{g (1 - a g)}\, d\theta^2 \nonumber\\
  {} & \quad + \Lambda_{\textrm{AdS$_5$}} (\theta) \left[d\hat{\phi}^+ + \frac{3 a (1 - a g)}{2 (1 + 5 a g) \sqrt{a \left(a + \frac{2}{g} \right)}} \, \hat{\rho}\, d\hat{x}^+ \right]^2 \nonumber\\
  {} & \quad + \frac{a \Big(4 - a g + 3 a g\, \textrm{cos} (2 \theta) \Big)\, \textrm{cos}^2 \theta}{2 g (1 - a g)^2} \left[d\hat{x}^- + \frac{6 a g\, \textrm{sin}^2 \theta}{4 - a g + 3 a g\, \textrm{cos} (2 \theta)} d\hat{\phi}^+ + \hat{V}_{\textrm{AdS$_5$}} (\theta)\, \hat{\rho}\, d\hat{x}^+ \right]^2 \, ,\label{eq:AdS5 LightCone Metric Poincare}
\end{align}
where
\be
  \hat{\phi}^+ \equiv \widetilde{\phi} - \frac{3 a (1 - a g)}{2 (1 + 5 a g) \sqrt{a \left(a + \frac{2}{g} \right)}}\, \Theta^+\, ,\quad \hat{x}^- \equiv x^- - \frac{3 a (1 - a g)}{2 (1 + 5 a g) \sqrt{a \left(a + \frac{2}{g} \right)}}\, \Theta^+\, .
\ee
Similar to \eqref{eq:AdS5 NHmetric standard form}, the near-horizon metric in the light-cone coordinates \eqref{eq:AdS5 LightCone Metric Poincare} can also be written in the standard form
\be\label{eq:AdS5 NHmetric LightCone standard form}
  ds^2 = f_0 (\theta) \left(- \hat{\rho}^2\, d\hat{x}^{+2} + \frac{d\hat{\rho}^2}{\hat{\rho}^2} \right) + f_\theta (\theta)\, d\theta^2 + \gamma_{ij} (\theta)\, \left(dx^i + k^i \hat{\rho}\, d\hat{x}^+ \right)\, \left(dx^j + k^j \hat{\rho}\, d\hat{x}^+ \right)\, ,
\ee
with $x^i \in \{\hat{\phi}^+,\, \hat{x}^- \}$, and $k_i$, $f_0 (\theta)$, $f_\theta (\theta)$ and $\gamma_{ij} (\theta)$ remain the same as \eqref{eq:AdS5 NHmetric standard form}. We will refer to these particular forms of the metric at various points in the coming subsection, where we recover the entropy from these data.

\subsection{Cardy Formula}

In this subsection, we first compute the central charges of the Virasoro algebras, and we specialize to the parameters $a = b$ of the non-extremal black holes for simplicity. The left central charge $c_L$ can be computed by applying the standard Kerr/CFT correspondence \cite{Guica:2008mu, Lu:2008jk} to the near-horizon metric \eqref{eq:AdS5 NH metric Global} in the global coordinates:
\be\label{eq:AdS5 VirasoroIntegral cL 1}
  \frac{1}{8 \pi G} \int_{\partial \Sigma} k_{\zeta_{(m)}} [\mathcal{L}_{\zeta_{(n)}} g,\, g] = - \frac{i}{12} c_L\, (m^3 + \alpha m)\, \delta_{m+n,\, 0}\, ,
\ee
with the mode expansion of a diffeomorphism generating vector $\zeta$ as
\be
  \zeta_{(n)} = - e^{- i n \widetilde{\phi}} \frac{\partial}{\partial \widetilde{\phi}} - i n r e^{- i n \widetilde{\phi}} \frac{\partial}{\partial \hat{r}}\, ,
\ee
while $k_\zeta$ stands for a 2-form defined for a general perturbation $h_{\mu\nu}$ around the background metric $g_{\mu\nu}$:
\begin{align}
  k_\zeta [h,\, g] & \equiv - \frac{1}{4} \epsilon_{\alpha\beta\mu\nu} \Big[\zeta^\nu D^\mu h - \zeta^\nu D_\sigma h^{\mu\sigma} + \zeta_\sigma D^\nu h^{\mu\sigma} + \frac{1}{2} h D^\nu \zeta^\mu - h^{\nu\sigma} D_\sigma \zeta^\mu \nonumber\\
  {} & \qquad\qquad\quad + \frac{1}{2} h^{\sigma\nu} (D^\mu \zeta_\sigma + D_\sigma \zeta^\mu)\Big] dx^\alpha \wedge dx^\beta\, ,
\end{align}
and $\mathcal{L}_\zeta$ denotes the Lie derivative with respect to $\zeta$:
\be
  \mathcal{L}_\zeta g_{\mu\nu} \equiv \zeta^\rho \partial_\rho g_{\mu\nu} + g_{\rho\nu} \partial_\mu \zeta^\rho + g_{\mu\rho} \partial_\nu \zeta^\rho\, .
\ee

More precisely, there are two slightly different approaches to compute the left central charge $c_L$ using the Kerr/CFT correspondence. One is to evaluate the integral \eqref{eq:AdS5 VirasoroIntegral cL 1}, where $g$ denotes the near-horizon metric of the near-extremal AdS$_5$ black hole in the global coordinates \eqref{eq:AdS5 NH metric Global}. An explicit evaluation of \eqref{eq:AdS5 VirasoroIntegral cL 1} shows that
\be\label{eq:AdS5 cL 1}
  c_L = \frac{9 \pi a^2}{g (1 - a g) (1 + 5 a g)}\, .
\ee
There is another approach of computing $c_L$  advanced in \cite{Matsuo:2010ut}, which consists in evaluating the following  integral expression
\be
  \frac{1}{8 \pi G_N} \int_{\partial \Sigma} k_{\xi_n} \left[\mathcal{L}_{\xi_m} \bar{g},\, \bar{g} \right] = \delta_{n+m,\, 0}\, n^3 \frac{c_L}{12}\, .
\ee
Let us highlight the subtle differences with respect to the expression \eqref{eq:AdS5 VirasoroIntegral cL 1}. First, the background metric $\bar{g}$ denotes the standard form of the near-horizon metric of the near-extremal AdS$_5$ black hole in the Poincar\'e coordinates \eqref{eq:AdS5 NHmetric standard form}. Note also that there is no reference to the linear in mode $n$ part of the Virasoro algebra. More precisely,
\be\label{eq:AdS5 VirasoroIntegral cL 2}
  c_L = \frac{6 \pi k_\phi}{G_N} \int_0^{\pi/2} d\theta\, \sqrt{\gamma (\theta)\, f_\theta (\theta)} = \frac{6 \pi k_\psi}{G_N} \int_0^{\pi/2} d\theta\, \sqrt{\gamma (\theta)\, f_\theta (\theta)}\, ,
\ee
where the factors $\gamma (\theta)$, $f_\theta (\theta)$, $k_\phi$ and $k_\psi$ are defined in \eqref{eq:AdS5 NHmetric standard form}. The explicit result of \eqref{eq:AdS5 VirasoroIntegral cL 2} is
\be\label{eq:AdS5 cL 2}
  c_L = \frac{9 \pi a^2}{g (1 - a g) (1 + 5 a g)}\, ,
\ee
which is exactly the same as the one from the first approach \eqref{eq:AdS5 cL 1}.

The right central charge $c_R$ can also be obtained in two different ways. The first approach is to compute the quasi-local charge \cite{Brown:1992br, Balasubramanian:1999re, Matsuo:2010ut} using the standard form of the near-horizon metric of the near-extremal AdS$_5$ black hole in the Poincar\'e coordinates \eqref{eq:AdS5 NHmetric standard form}, which is given by the integral
\be\label{eq:AdS5 VirasoroIntegral cR 1}
  \frac{c_R}{12} = \frac{1}{8 \pi G_N} \int d\hat{\phi}\, d\hat{\psi}\, d\theta\, \frac{k_i k_j \gamma_{ij} (\theta)\, \sqrt{\gamma (\theta)\, f_\theta (\theta)}}{2 \Lambda_0 f_0 (\theta)}\, ,
\ee
where the factors $\gamma (\theta)$, $\gamma_{ij} (\theta)$, $f_0 (\theta)$, $f_\theta (\theta)$, $k_i$ are defined in \eqref{eq:AdS5 NHmetric standard form}, and the parameter $\Lambda_0$ denotes a UV cutoff in $r$. The explicit result of \eqref{eq:AdS5 VirasoroIntegral cR 1} is
\begin{align}
  c_R = \frac{27 \pi a^{\frac{5}{2}} \sqrt{a + \frac{2}{g}} }{2 g^2 (1 - a g)^2 (1 + 5 a g) \Lambda_0}\, ,
\end{align}
but it contains an  arbitrary cutoff $\Lambda_0$, which makes this approach not very attractive.

To explicitly compute the right central charge $c_R$ independent of a cutoff, a new near-horizon limit was introduced in \cite{Matsuo:2010ut}, which has also been applied to near-extremal AdS$_5$ black hole solutions in the previous subsection. A scale-covariant right central charge $c_R^{(cov)}$ can be computed from the standard form of the near-horizon metric of the near-extremal AdS$_5$ black hole in the light-cone coordinates \eqref{eq:AdS5 NHmetric LightCone standard form} by using the following integral \cite{Matsuo:2010ut}
\be\label{eq:AdS5 VirasoroIntegral cR 2}
  c_R^{(cov)} = 6 \pi k \lambda \int_0^{\pi/2} d\theta\, \sqrt{\gamma (\theta)\, f_\theta (\theta)}\, ,
\ee
where the factors $\gamma (\theta)$, $f_\theta (\theta)$, $k_\phi$ and $k_\psi$ are defined in \eqref{eq:AdS5 NHmetric LightCone standard form}. The explicit result of \label{eq:AdS5 VirasoroIntegral cR 2} is
\be\label{eq:AdS5 cR 2}
  c_R^{(cov)} = \frac{9 \pi a^2 \lambda}{g (1 - a g) (1 + 5 a g)}\, .
\ee
Like in \cite{Matsuo:2010ut}, we can define a scale-invariant right central charge $c_R \equiv c_R^{(cov)} / \lambda$, which in this case is
\be
  c_R = \frac{9 \pi a^2}{g (1 - a g) (1 + 5 a g)}\, .
\ee
It is exactly the same as the left central charge \eqref{eq:AdS5 cL 1} \eqref{eq:AdS5 cL 2} from two different approaches.

Besides the central charges $c_{L}=c_{R}$ obtained in this subsection, we still need the Frolov-Thorne temperatures to compute the near-extremal black hole entropy. The BPS AdS$_5$ black hole case has been discussed in \cite{David:2020ems}, for which $T_L \neq 0$ and $T_R = 0$. For the near-extremal AdS$_5$ black hole, $T_L$ remains the same as the BPS value \cite{David:2020ems}, while $T_R$ has a nonvanishing value proportional to the physical Hawking temperature $T_H$.

To compute the Frolov-Thorne temperatures for the near-extremal case, we first consider the general non-extremal case and then take the near-extremal limit. The non-extremal $T_L$ and $T_R$ can be obtained using the hidden conformal symmetry approach. The basic idea of this approach is to find a set of near-horizon conformal coordinates \cite{Castro:2010fd}, under which the wave equation of an uncharged massless scalar field is given by the quadratic Casimir of the $SU(2, \mathbb{R})$ Lie algebra. In particular, the wave equation in the non-extremal AdS$_5$ black hole background \eqref{eq:AdS5metric} was already discussed in \cite{Wu:2009ug}, and we can apply the technique similar to the AdS$_4$ black hole case discussed in \cite{Chen:2010bh} to find the AdS$_5$ near-horizon conformal coordinates and consequently the Frolov-Thorne temperatures for non-extremal AdS$_5$ black holes. The results for $b = a$ are
\begin{align}
\begin{split}\label{eq:Frolov Thorne TL TR}
  T_L & = \frac{k \Big[ a^2 (a^2 + q) r_+^2 + \left(a^4 + r_+^4 + a^2 (q + 4\, r_+^2) \right) r_*^2 - r_+^2 r_*^4 \Big]}{4 \pi a \Xi_a (r_+ + r_*) \Big[ (a^2 + q) (a^2 + r_+^2) + (a^2 + q + r_+^2) r_*^2 \Big]}\, ,\\
  T_R & = \frac{- k (r_+ - r_*) (a^4 + a^2 q - r_+^2 r_*^2)}{4 \pi a \Xi_a \Big[ (a^2 + q) (a^2 + r_+^2) + (a^2 + q + r_+^2) r_*^2 \Big]}\, ,
\end{split}
\end{align}
where $k$ and $r_*$ can be obtained by expanding $\Delta_r$ to the quadratic order in $(r - r_+)$, similar to \cite{Chen:2010bh}. More explicitly, for the general non-extremal AdS$_5$ black hole case with $b = a$, the explicit expressions of $k$ and $r_*$ are
\begin{align}
\begin{split}
  k & = 1 + 2 a^2 g^2 + 6 g^2 r_+^2 + \frac{3 (a^2 + q)^2}{r_+^4}\, ,\\
  r_* & = \frac{2 (3 - 2 a g) (a^2 + q)^2 r_+ - 2 (1 + 2 a^2 g^2) r_+^5 - 4 a g^3 r_+^7}{(1 - a g) \Big[3 (a^2 + q)^2 + (1 + 2 a^2 g^2) r_+^4 + 6 g^2 r_+^6 \Big]}\, ,
\end{split}
\end{align}
where $r_* \to r_+$ in the BPS limit.

For the near-extremal case, we impose the conditions \eqref{eq:AdS5 NearBPScond} and $q = m_0 / (1 + a g + b g)$, and simplify the Frolov-Thorne temperatures \eqref{eq:Frolov Thorne TL TR} as follows
\be
  T_L = \frac{(1 + 5 a g) \sqrt{a \left(a + \frac{2}{g} \right)}}{3 \pi a (1 - a g)}\, ,\quad T_R = \frac{3 - a g}{3 g (1 - a g)}\, T_H\, ,
\ee
where $T_H$ is the Hawking temperature given by \eqref{eq:temperature}. These values of $T_{L, R}$ are consistent with \eqref{eq:DefTL} \eqref{eq:DefTR} expected from the boundary $\mathcal{N}=4$ SYM. With the explicit results of the central charges $c_{L, R}$ and the Frolov-Thorne temperatures $T_{L, R}$, the near-extremal AdS$_5$ black hole entropy can be obtained by using the Cardy formula:
\be\label{eq:NearExtSBHfromCardy}
  S_\pm = \frac{\pi^2}{3} c_L T_L \pm \frac{\pi^2}{3} c_R T_R = \frac{\pi^2 a \sqrt{2 a + a^2 g}}{g^{3/2} (1 - a g)^2} \pm \frac{a^2 (3 - a g) \pi^3}{g^2 (1 - a g)^2 (1 + 5 a g)} T_H\, ,
\ee
which justifies the definitions in \eqref{eq:S+ in c and T}, and reproduces the result \eqref{eq:C over T from bdy} obtained from the boundary $\mathcal{N}=4$ SYM and \eqref{eq:EntropyEnergyOuterHorizon} \eqref{eq:C over T from gravity} from the gravity solution.

This analysis solidifies the decomposition of the $\mathcal{N}=4$ SYM results into the left and the right sectors, which becomes manifest from the near-horizon CFT$_2$.  As discussed in \cite{David:2020jhp}, the same approach can be applied to other asymptotically AdS black holes in various dimensions. Besides the near-horizon limit \eqref{eq:NearHorizonLimit} and the near-extremal Kerr/CFT correspondence discussed in this subsection, one can also directly study the AdS$_2$ in the near-horizon region and apply the AdS$_2$/CFT$_1$ correspondence to the near-extremal black holes \cite{Castro:2009jf, Castro:2018ffi}.

\section{Hawking Radiation Rate}\label{sec:HawkingRadiation}

Having motivated the appearance of an effective CFT$_2$ from the point of view of ${\cal N}=4$ SYM and from the near-horizon geometry, let us discuss, following closely \cite{Callan:1996dv,Das:1996ug,Horowitz:1996fn}, one final implication -- the rate of Hawking radiation. The key observation is that any CFT$_2$ comes naturally equipped with the ``momentum'' operator which breaks the symmetry between the left- and the right-moving  sectors. For the near-extremal case, we consider the following physical limit of the left-moving and the right-moving modes:
\be
  N_L = N_L^* + \delta N_L\, ,\quad N_R = \delta N_R\, ,\quad \textrm{with}\quad \delta N_L = \delta N_R \ll N_L^*\, .
\ee
Assuming that the right-moving modes form a canonical ensemble, then the partition function of the right sector is

\be
  Z_R = \sum_{N_R} q^{N_R} d(N_R) = \sum_{N_R} q^{N_R} e^{S_R} = \sum_{N_R} q^{N_R} e^{2 \pi \sqrt{c_R N_R / 6}}\, .
\ee
Performing a saddle-point approximation with respect to $N_R$, we obtain the number of right-moving modes:
 \be\label{eq:log(q)}
  \delta N_R = N_R = q \frac{\partial}{\partial q} \textrm{log} Z \approx \frac{\pi^2 c_R}{6\, \left(\textrm{log} (q) \right)^2}\, ,\quad \textrm{with}\quad \textrm{log} (q) < 0\, .
\ee 
The occupation number in the right sector is given by the Bose-Einstein statistics:
\be\label{eq:Rel q and T}
  \rho_R (k_0) = \frac{q^n}{1 - q^n} = \frac{e^{- \frac{k_0}{T_R}}}{1- e^{- \frac{k_0}{T_R}}}\, ,
\ee
where $n$ denotes the momentum quantum number of a mode, that moves in the presence of a time circle for the near-horizon region of AdS$_5$ black holes. From \eqref{eq:log(q)} and \eqref{eq:Rel q and T} we obtain
\be
  T_R = \frac{k_0}{\pi n} \sqrt{6\, \frac{\delta N_R}{c_R}} = \frac{1}{\pi} \sqrt{6\, \frac{\delta N_R}{c_R}} = \frac{1}{\pi} \sqrt{\frac{6\, N_R}{c_R}}\, ,
\ee
where we used $k_0 = n$. A similar expression holds for $T_L$. Since we consider the limit $k_0 \sim T_R \ll T_L$, the occupation number in the left sector is
 \be\label{eq: rhoL approx}
  \rho_L (k_0) = \frac{e^{- \frac{k_0}{T_L}}}{1- e^{- \frac{k_0}{T_L}}} \approx \frac{T_L}{k_0} = \frac{1}{\pi k_0} \sqrt{\frac{6\, N_L}{c_L}}\, .
\ee

As discussed in \cite{Callan:1996dv}, the Hawking radiation for 5d asymptotically flat near-extremal black holes can be viewed as a scattering process between the left-moving and the right-moving modes in the D1-D5 CFT. Based on our discussion, it is tempting to interpret the Hawking radiation in asymptotically AdS$_5$ near-extremal black holes as a scattering process between the left-moving and the right-moving modes of the open strings living on a stack of black D3-branes. Near-extremal black D3-branes have been studied in \cite{Gubser:1996de} and a derivation of Hawking radiation from black D3-brane dynamics was provided. The Hawking radiation rate for asymptotically flat black holes has also been studied using Kerr/CFT in \cite{Cvetic:2009jn}. Now we are able to find the Hawking radiation from the dual $\mathcal{N}=4$ SYM on the boundary, as expected from the AdS/CFT correspondence \cite{Maldacena:1997re}. Following \cite{Callan:1996dv} one can evaluate the radiation rate as
\be
  d\Gamma \sim \frac{d^4 k}{k_0} \frac{1}{p_0^L\, p_0^R} |\mathcal{A}|^2\, c_L\, \rho_L (k_0)\, \rho_R (k_0)\, ,
\ee
where $\mathcal{A}$ denotes the disc amplitude of strings, and the central charge $c_L$ accounts for the degrees of freedom for a given momentum $n$. Due to \eqref{eq:Rel q and T} and \eqref{eq: rhoL approx}, we see that
\be
  c_L\, \rho_L (k_0) \sim S_L \propto (\textrm{horizon area})\, ,
\ee
and consequently,
\be
  d\Gamma \sim (\textrm{horizon area})\cdot \frac{e^{- \frac{k_0}{T_R}}}{1- e^{- \frac{k_0}{T_R}}} d^4 k\, .
\ee
Hence, the radiation rate is proportial to the area of the black hole, and has a spectrum obeying the Bose-Einstein statistics with the Hawking temperature $T_H \sim T_R$. According to this picture, the scattering of modes is unitary, hence there is no information loss during the Hawking radiation process. Clearly, global aspects of asymptotically AdS$_5$ spacetimes play an important role in tracking Hawking radiation. In particular, it is clear that radiation eventually bounces at the asymptotic boundary of the spacetime and returns to the center. Therefore, the description we provide here should be understood exclusively as an approximate description valid in the region close to the horizon in the sense we have presented  and should not be construed to address the late-time aspects of black hole evaporation.

\section{Discussions}\label{sec:Discussions}

In this paper, we studied the thermodynamics of electrically charged rotating near-extremal AdS$_5$ black holes both from the gravity and from the dual $\mathcal{N}=4$ SYM. We started from the BPS AdS$_5$ black holes previously studied in the literature \cite{Cabo-Bizet:2018ehj, Choi:2018hmj, Benini:2018ywd} and considered perturbations around this configuration by introducing  a small temperature. We carefully distinguished the outer and the inner horizons and found first laws similar to asymptotically flat black holes \cite{Castro:2012av,Cvetic:2018dqf}. We reinterpreted these laws in terms of the left- and right-moving sectors of an effective CFT$_2$ arising directly from ${\cal N}=4$ SYM. We corroborated the existence of the effective CFT$_2$ by showing its geometric appearance as the set of asymptotic symmetries of a certain near-horizon limit. Armed with this effective CFT$_2$, we provided a path for understanding aspects of the Hawking radiation rate in near-extremal AdS$_5$ black holes similar to the asymptotically flat case \cite{Callan:1996dv,Das:1996ug,Das:1996wn}.

Our results suggest that the universality of black hole quantum dynamics in AdS$_5$ can be related holographically to the existence of a local AdS$_3$ region near the horizon \cite{Johnstone:2013eg, Goldstein:2019gpz, David:2020ems}. It would be precisely because of this region that the effective  CFT$_2$ description arises and can be, in principle, understood from the relation between some 4d superconformal algebra and its corresponding 2d CFT. It is also plausible that similar near-extremal relations and effective CFT$_2$ interpretations exist for asymptotically AdS black holes in other dimensions \cite{David:2020jhp}, but each case deserves more detailed analysis.

\section*{\bf Acknowledgments}
We would like to thank M. Cvetic, M. David, T. Faulkner, A. Gonz\'alez Lezcano, A. Herderschee, S. Jeong, C. Keeler, J. Kim, F. Larsen, J. T. Liu, V. Martin, G. Moore, R. Myers, W. Song, A. Svesko, J. Yagi and X. Zhang for discussions. This  work was supported in part by the U.S. Department of Energy under grant DE-SC0007859. J.N was also supported by a Van Loo Postdoctoral Fellowship.

\bibliographystyle{utphys}
\bibliography{HawkingRadiation}
\end{document}